\documentstyle[12pt]{article}
\begin{document}
\hfuzz=1pt
\setlength{\textheight}{8.5in}
\setlength{\topmargin}{0in}
\begin{center}
\Large {\bf  Maximal violation of Bell's inequality 
in the case of real experiments}
\\  \vspace{.75in}
\large {M. Ardehali}\footnote[1]
{email address:ardehali@apexmail.com}
\\ \vspace{.3in}
Research Laboratories,
NEC Corporation,\\
Sagamihara,
Kanagawa 229
Japan

\end{center}
\vspace{.20in}

\begin{abstract}
Einstein's locality is invoked to derive a correlation inequality.
In the case of 
ideal experiments, this inequality
is equivalent to Bell's original inequality of $1965$ which, as
is well known, is violated by a maximum factor of $1.5$.
The crucial point
is that even in the case of
real experiments where polarizers and detectors are non-ideal,
the present inequality
is violated by a factor of $1.5$, whereas
previous inequalities such as
Clauser-Horne-Shimony-Holt inequality of $1969$
and Clauser-Horne inequality of $1974$
are violated by a factor of $\sqrt 2$.
The larger magnitude of violation 
can be of importance for the experimental
test of locality.
Moreover, the  supplementary assumption used to derive this
inequality is weaker than
Garuccio-Rapisarda assumption.
Thus an experiment based on this inequality
refutes a larger family
of hidden variable theories than an experiment based on 
Garuccio-Rapisarda inequality.
\end {abstract}
\pagebreak

\begin{center}
\Large  {\bf I. Introduction}
\end{center}

Local realism is a philosophical view which holds that external
reality exists and has local properties.
Quantum mechanics vehemently denies that
such a world view has any meaning for physical systems
because local realism
assigns simultaneous values to non-commuting observables.
In $1965$ Bell \cite{1} showed that the assumption  of local realism, as
postulated by Einstein, Podolsky, and Rosen (EPR) \cite{2},
leads to some
constraints on the statistics of two spatially separated particles.
These constraints, which are collectively known as Bell inequalities,
are sometimes grossly violated by quantum mechanics. The violation of
Bell inequalities therefore indicate that local realism is not only
philosophically but also numerically
incompatible with quantum mechanics.
Bell's theorem is of paramount importance for undersanding the
foundations of quantum mechanics because it rigorously formulates
EPR's assumption of locality and shows that
all realistic interpretations of quantum mechanics must be nonlocal.

Bell's original argument, however,
can not be experimentally tested
because it relies on perfect correlation of the spin of the two
particles \cite {3}. Faced with this problem,
Clauser-Horne-Shimony-Holt (CHSH) \cite{4},
Freedman-Clauser (FC) \cite{5}, and Clauser-Horne (CH) \cite{6}
derived correlation
inequalities for
systems which do not achieve $100\%$ correlation,
but which do achieve a necessary minimum correlation. 
An Experiment based on CHSH, or FC,
or CH inequality utilizes one-channel
polarizers in which the dichotomic choice is between the detection of
the photon and its lack of detection. A better experiment is
one in which
a truly binary choice
is made between the ordinary and the extraordinary rays [7-10].
In this paper, we derive a correlation inequality
for two-channel polarizer systems and we show that 
quantum mechanics violates this inequality 
by a factor of $1.5$,
whereas it violates the previous inequalities [4-10] by a factor of
$\sqrt 2$. 
Thus the magnitude of
violation of the inequality derived in this paper
is approximately $20.7\%$ larger than
the magnitude of violation of previous
inequalities [4-10]. Moreover, we show that
the present inequality
requires the
measurement of only three detection probabilities, whereas
CH (or CHSH) inequality
requires the measurements
of five detection probabilities.
Thus the present inequality can 
be used to test locality more simply
than CH (or CHSH) inequality.

\begin{center}
\Large  {\bf II. Experiments with pairs of atomic photons}
\end{center}

We start by considering Bohm's \cite{11} version of EPR experiment
in which an unstable source emits pairs of photons in
a cascade from state $J=1$ to $J=0$
(see Fig. 1 of Ref. 10).
The source is viewed by two
apparatuses.
The first (second) apparatus consists of a polarizer
$P_1 \left(P_2 \right)$
set at angle $\mbox{\boldmath $m$} \left(
\mbox{\boldmath $n$} \right)$,
and two detectors
$D_{1}^{\,\pm} \left (D_{2}^{\,\pm} \right)$
put along the ordinary and the extraordinary beams.
During a period of time $T$ 
while the polarizers are set along axes
$\mbox{\boldmath $m$}$ and 
$\mbox{\boldmath $n$}$, the source emits, say, $N$ pairs of
photons.

Let $N^{\,\pm\,\pm}\left(\mbox{\boldmath $m,n$}\right)$
be the number of simultaneous counts from 
detectors $D_{1}^{\pm}$ and $D_{2}^{\pm}$,
$N^{\,\pm\,0}\left(\mbox{\boldmath $m,n$}\right)$
the number of counts when 
detectors $D_{1}^{\pm}$ are
triggered but detectors $D_{2}^{\pm}$ are not triggered,
$N^{\,0 \,  \pm}\left(\mbox{\boldmath $m,n$}\right)$
the number of counts when detectors
$D_{2}^{\pm}$ are
triggered but $D_{1}^{\pm}$ are not triggered, and finally
$N^{\,0\,0}\left(\mbox{\boldmath $m,n$}\right)$
the number of photons that are emitted by the source but not detected
by either 
$D_{1}^{\pm}$ or
$D_{2}^{\pm}$.
If the time $T$ is sufficiently
long, then the ensemble probabilities
are defined as
\begin{eqnarray}{\nonumber}
&&p^{\;\pm\;\pm} \left(\mbox{\boldmath $m,n$} \right)=
\frac{N^{\;\pm\;\pm} \left(\mbox{\boldmath $m,n$} \right)}{N}, \qquad
p^{\;\pm\; 0} \left(\mbox{\boldmath $m,n$} \right)=
\frac{N^{\;\pm \;  0} \left(\mbox{\boldmath $m,n$} \right)}{N}, \\
&&p^{\;0\;\pm} \left(\mbox{\boldmath $m,n$} \right)=
\frac{N^{\;0\;\pm} \left(\mbox{\boldmath $m,n$} \right)}{N}, \qquad
p^{\;0\; 0} \left(\mbox{\boldmath $m,n$} \right)=
\frac{N^{\;0\;0} \left(\mbox{\boldmath $m,n$} \right)}{N}.
\end{eqnarray}
Similarly, we let
$N^{\,\pm}\left(\mbox{\boldmath $m$}\right)$ 
[$N^{\,\pm}\left(\mbox{\boldmath $n$}\right)$] be
the number of counts from detectors
$D_1^\pm$ [$D_2^\pm$], and
$N^{\,0}\left(\mbox{\boldmath $m$}\right)$
[$N^{\,0}\left(\mbox{\boldmath $n$}\right)$]
the number of photons that are emitted by the source but not detected
by $D_1^\pm$
[$D_2^\pm$].
Again if the time $T$ is sufficiently
long, then the ensemble probabilities 
are defined as
\begin{eqnarray}{\nonumber}
&&p^{\;\pm}(\mbox{\boldmath $m$})=
\frac{N^{\;\pm}(\mbox{\boldmath $m$})}{N}, \qquad
p^{\;0}(\mbox{\boldmath $m$})=
\frac{N^{\;0}(\mbox{\boldmath $m$})}{N}, \\ 
&&p^{\;\pm}(\mbox{\boldmath $n$})=
\frac{N^{\;\pm}(\mbox{\boldmath $n$})}{N}, \qquad
p^{\;0}(\mbox{\boldmath $n$})=
\frac{N^{\;0}(\mbox{\boldmath $n$})}{N}.
\end{eqnarray}
It is important to emphasize that in real experiments, 
due to imperfection of
polarizers and detectors,
$p^{\;\pm \; 0} \left(\mbox{\boldmath $m,n$} \right)$,
$p^{\;0 \; \pm} \left(\mbox{\boldmath $m,n$} \right)$, and
$p^{\;0 \; 0} \left(\mbox{\boldmath $m,n$} \right)$
are non-zero; in fact in experiments which
are feasible with present technology, these probabilities are
much larger than
$p^{\;\pm \; \pm} \left(\mbox{\boldmath $m,n$} \right)$ 
{\large(}similarly $p^{\;0}(\mbox{\boldmath $m$})$[
$p^{\;0}(\mbox{\boldmath $n$})$] are much larger than
$p^{\;\pm}(\mbox{\boldmath $m$})$[
$p^{\;\pm}(\mbox{\boldmath $n$})$]{\large )}. Since 
$p^{\;\pm \; 0} \left(\mbox{\boldmath $m,n$} \right)$,
$p^{\;0 \; \pm} \left(\mbox{\boldmath $m,n$} \right)$,
$p^{\;0 \; 0} \left(\mbox{\boldmath $m,n$} \right)$,
$p^{\;0}(\mbox{\boldmath $m$})$, and
$p^{\;0}(\mbox{\boldmath $n$})$
can not be measured in actual experiments, it is crucial that they do
not appear in any
correlation inequality that is used to test locality.

We now consider a particular pair of photons and specify its
state with a parameter $\lambda$. Following Bell, we do not 
impose any restriction on the complexity of $\lambda$. 
``It is
a matter of indifference
in the following whether $\lambda$ denotes a single variable or
a set, or even a set of functions, and whether the variables are 
discrete or continuous \cite{1}.''

The ensemble probabilities
in Eqs. $(1)$ and $(2)$ are defined as
\begin{eqnarray} {\nonumber}
p^{\:\pm\:\pm}(\mbox{\boldmath $a,b$}) &=&
\int p\,(\lambda)\, p^{\;\pm}(\mbox{\boldmath 
$a$} \mid \lambda) \, p^{\;\pm}(\mbox{\boldmath $b$}
\mid \lambda,\mbox{\boldmath $a$}), \\ \nonumber
p^{\:\pm}(\mbox{\boldmath $a$}) &=&
\int p \, (\lambda) \, p^{\;\pm}(\mbox{\boldmath 
$a$} \mid \lambda), \\ 
p^{\:\pm}(\mbox{\boldmath $b$}) &=&
\int p \, (\lambda)\, p^{\;\pm}(\mbox{\boldmath 
$b$} \mid \lambda).
\end{eqnarray}
Equations $(3)$ may be stated in physical terms: The ensemble
probability for detection of photons by
detectors $D^{\;\pm}_{\; 1}$ and $D^{\;\pm}_{\;2}$
[that is $p^{\;\pm\;\pm}(\mbox{\boldmath $a,b$})$]
is equal to the sum or integral of the probability
that the emission is
in the state $\lambda$ [that is $p(\lambda)$], times the conditional
probability that if the emission is in the state $\lambda$,
then a count is triggered by the first detector $D^{\;\pm}_{1}$
[that is $p^{\;\pm}(\mbox{\boldmath $a$}
\mid \lambda)$],
times the conditional probability that 
if the emission is in the state 
$\lambda$ and if the first polarizer is set along axis $\boldmath a$,
then a count is triggered from the second detector $D^{\;\pm}_{2}$
[that is $p^{\;\pm}(\mbox{\boldmath $b$}
\mid \lambda,\mbox{\boldmath $a$})$].
Similarly the ensemble probability for detection of photons by
detector $D^{\;\pm}_{\;1} \left(D^{\;\pm}_{\;2} \right )$
{\large [} that is $p^{\;\pm}(\mbox{\boldmath $a$}) \left
[p^{\;\pm}(\mbox{\boldmath $b$}) \right]$ {\large ]}
is equal to the sum or integral of the probability that the photon
is in the state $\lambda$ [that is $p(\lambda)$], times the
conditional probability that if the
photon is in the state $\lambda$,
then a count is triggered by
detector $D^{\;\pm}_{1} \left(D^{\;\pm}_{2} \right )$
{\large[} that is $p^{\;\pm}(\mbox{\boldmath $a$}
\mid \lambda) \left [p^{\;\pm}(\mbox{\boldmath $b$} \mid 
\lambda ) \right ]$ {\large]}.
Note that Eqs. $(1)$, $(2)$, and $(3)$ are quite general and follow
from the standard rules of probability theory.
No assumption has yet been made that is not satisfied 
by quantum mechanics.

Hereafter, we
focus our attention only on those theories that satisfy
EPR criterion of locality: ``Since at the time of measurement the
two systems no longer interact, no real change can take place in the
second system in consequence of anything that may be done to first
system
\cite {2}''. EPR's criterion of locality can be translated into
the following mathematical equation:
\begin{equation}
p^{\;\pm}(\mbox{\boldmath $b$} \mid \lambda,
\mbox{\boldmath $a$})=
p^{\;\pm}(\mbox{\boldmath $b$} \mid \lambda).
\end{equation}
Equation $(4)$ is the hall mark of local realism.
It is the most general form of locality that accounts
for correlations subject only to the requirement that a count
triggered by the second detector does not depend on
the orientation of the first polarizer. The assumption
of locality, i.e., Eq. $(4)$, is
quite natural since the two photons are spatially separated so that
the orientation of the first polarizer should not influence the
measurement carried out on the second photon.

\begin{center}
\Large  {\bf III. Bell's inequality}
\end{center}

In the following we show that 
equation $(4)$ leads to validity of an equality
that is sometimes grossly violated by
the quantum mechanical predictions in the case of real experiments.
First we need to prove the following algebraic theorem.

{\it Theorem:} Given ten non-negative real numbers
$x_{1}^{+}$, $x_{1}^{-}$, $x_{2}^{+}$, $x_{2}^{-}$,
$y_{1}^{+}$, $y_{1}^{-}$, $y_{2}^{+}$, $y_{2}^{-}$, $U$ and $V$
such that
$x_{1}^{+}, x_{1}^{-},
x_{2}^{+}, x_{2}^{-} \leq U$,
and
$y_{1}^{+}, y_{1}^{-},
y_{2}^{+}, y_{2}^{-} \leq V$,
then the following inequality always holds:
\begin{eqnarray}{\nonumber}
Z &=& x_{1}^{+}y_{1}^{+}
+x_{1}^{-}y_{1}^{-}
-x_{1}^{+}y_{1}^{-}
-x_{1}^{-}y_{1}^{+}
+y_{2}^{+}x_{1}^{+}
+y_{2}^{-}x_{1}^{-} \\ \nonumber
&-&y_{2}^{+}x_{1}^{-}
-y_{2}^{-}x_{1}^{+}
-y_{1}^{+}x_{2}^{+}
-y_{1}^{-}x_{2}^{-}
+y_{1}^{+}x_{2}^{-}
+y_{1}^{-}x_{2}^{+}
+2x_{2}^{+}y_{2}^{+} \\
&+&2x_{2}^{-}y_{2}^{-}
-Vx_{2}^{+}-Vx_{2}^{-}
-Uy_{2}^{+}-Uy_{2}^{-} - UV \le 0.
\end{eqnarray}
{\it Proof}:
Calling $A=y_{1}^{+}-y_{1}^{-}$, we write the function $Z$ as
\begin{eqnarray} {\nonumber}
Z&=&
x_{2}^{+}
\left(2y_{2}^{+} - A - V \right )
+ x_{2}^{-}
\left(2y_{2}^{-} + A - V \right ) \\
&+&\left( x_{1}^{+} - x_{1}^{-} \right )
\left(A + y_{2}^{+}-y_{2}^{-} \right )
- U y_{2}^{+} -  U y_{2}^{-}
-UV.
\end{eqnarray}

\noindent We consider the following eight cases:
\\
(1) First assume
$\left \{
\begin{array}{c}
2y_{2}^{+} - A  - V \le 0,\\
2y_{2}^{-} + A - V \le 0,\\
A + y_{2}^{+}-y_{2}^{-}\le 0.
\end{array} \right.$
\vspace{0.4 cm}

\noindent The function $Z$ is maximized if
$x_{2}^{+}=0, x_{2}^{-}=0$, and
$ x_{1}^{+} - x_{1}^{-} =-U$. Thus
\begin{eqnarray} {\nonumber}
Z &\le&
-U \left(A +y_{2}^{+} - y_{2}^{-} \right )
-U y_{2}^{+} -  U y_{2}^{-}
-UV \\
&=&-U\left(A + 2y_{2}^{+} + V\right ).
\end{eqnarray}
Since $V \ge A$ and $y_{2}^{+} \ge 0$, $Z \le 0$.
\vspace{0.7 cm}
\\
(2) Next assume
$\left \{
\begin{array}{c}
2y_{2}^{+} - A - V > 0,\\
2y_{2}^{-} + A - V \le 0,\\
A + y_{2}^{+}-y_{2}^{-}\le 0.
\end{array} \right.$
\vspace{0.4 cm}

\noindent The function $Z$ is maximized if
$x_{2}^{+}=U, x_{2}^{-}=0$, and
$ x_{1}^{+} - x_{1}^{-} =-U$. Thus
\begin{eqnarray} {\nonumber}
Z &\le&
U
\left(2y_{2}^{+} - A - V \right ) -
U \left(A + y_{2}^{+}-y_{2}^{-} \right )
- U y_{2}^{+} -  U y_{2}^{-}
-UV \\
&=&-2U\left(V+A\right ).
\end{eqnarray}
Since $V \ge A$, $Z \le 0$.
\vspace{0.7 cm}
\\
(3) Next assume
$\left \{
\begin{array}{c}
2y_{2}^{+} - A - V \le 0,\\
2y_{2}^{-} + A - V > 0,\\
A + y_{2}^{+}-y_{2}^{-}\le 0.
\end{array} \right.$
\vspace{0.4 cm}

\noindent The function $Z$ is maximized if
$x_{2}^{+}=0, x_{2}^{-}=U$, and
$ x_{1}^{+} - x_{1}^{-} =-U$. Thus
\begin{eqnarray} {\nonumber}
Z &\le&
U
\left(2y_{2}^{-} + A - V \right ) -
U \left(A + y_{2}^{+}-y_{2}^{-} \right )
- U y_{2}^{+} -  U y_{2}^{-}
-UV \\
&=&-2U\left (V -y_{2}^{-} +y_{2}^{+}\right).
\end{eqnarray}
Since $V \ge  y_{2}^{-}$, and $y_{2}^{+}\ge 0$, $Z \le 0$.
\vspace{0.7 cm}
\\
(4) Next assume
$\left \{
\begin{array}{c}
2y_{2}^{+} - A - V \le 0,\\
2y_{2}^{-} + A - V \le 0,\\
A + y_{2}^{+}-y_{2}^{-} > 0.
\end{array} \right.$
\vspace{0.4 cm}

\noindent The function $Z$ is maximized if
$x_{2}^{+}=0, x_{2}^{-}=0$, and
$ x_{1}^{+} - x_{1}^{-} =U$. Thus
\begin{eqnarray} {\nonumber}
Z &\le&
U \left(A + y_{2}^{+}-y_{2}^{-} \right )
- U y_{2}^{+} -  U y_{2}^{-}
-UV \\
&=&-U\left(-A + 2y_{2}^{-} + V\right ).
\end{eqnarray}
Since $V \ge -A$ and $y_{2}^{-} \ge 0$, $Z \le 0$.
\vspace{0.7 cm}
\\
(5) Next assume
$\left \{
\begin{array}{c}
2y_{2}^{+} - A - V > 0,\\
2y_{2}^{-} + A - V > 0,\\
A + y_{2}^{+}-y_{2}^{-}\le 0.
\end{array} \right.$
\vspace{0.4 cm}

\noindent The function $Z$ is maximized if
$x_{2}^{+}=U, x_{2}^{-}=U$, and
$ x_{1}^{+} - x_{1}^{-} =-U$. Thus
\begin{eqnarray} {\nonumber}
Z &\le&
U
\left(2y_{2}^{+} - A - V \right )
+U
\left(2y_{2}^{-} + A - V \right )
-U \left(A + y_{2}^{+}-y_{2}^{-} \right )\\ \nonumber
&-& U y_{2}^{+} -  U y_{2}^{-}
-UV \\ 
&=&-U\left(-2y_{2}^{-} +A +3V \right ).
\end{eqnarray}
Since $V \ge A$ and $V \ge y_{2}^{-}$, $Z \le 0$.
\vspace{0.7 cm}
\\
(6) Next assume
$\left \{
\begin{array}{c}
2y_{2}^{+} - A - V > 0,\\
2y_{2}^{-} + A - V \le 0,\\
A + y_{2}^{+}-y_{2}^{-} > 0.
\end{array} \right.$
\vspace{0.4 cm}

\noindent The function $Z$ is maximized if
$x_{2}^{+}=U, x_{2}^{-}=0$, and
$ x_{1}^{+} - x_{1}^{-} =U$. Thus
\begin{eqnarray} {\nonumber}
Z &\le&
U\left(2y_{2}^{+} - A - V \right )+
U \left(A + y_{2}^{+}-y_{2}^{-} \right )
- U y_{2}^{+} -  U y_{2}^{-}
-UV \\ 
&=&-2U\left(-y_{2}^{+}+y_{2}^{-} + V \right ).
\end{eqnarray}
Since $V \ge y_{2}^{+}$, and $y_{2}^{-} \ge 0$ $Z \le 0$.
\vspace{0.7 cm}
\\
(7) Next assume
$\left \{
\begin{array}{c}
2y_{2}^{+} - A - V \le 0,\\
2y_{2}^{-} + A - V > 0,\\
A + y_{2}^{+}-y_{2}^{-} > 0.
\end{array} \right.$
\vspace{0.4 cm}

\noindent The function $Z$ is maximized if
$x_{2}^{+}=0, x_{2}^{-}=U$, and
$ x_{1}^{+} - x_{1}^{-} = U$. Thus
\begin{eqnarray} {\nonumber}
Z &\le&
U
\left(2y_{2}^{-} + A - V \right )
+U \left(A + y_{2}^{+}-y_{2}^{-} \right )
- U y_{2}^{+} -  U y_{2}^{-}
-UV \\ 
&=&-2U\left( -A + V \right ).
\end{eqnarray}
Since $V \ge -A$, $Z \le 0$.
\vspace{0.7 cm}
\\
(8) Finally assume
$\left \{
\begin{array}{c}
-2y_{2}^{+} + A + V > 0,\\
-2y_{2}^{-} - A + V > 0,\\
A + y_{2}^{+}-y_{2}^{-} > 0.
\end{array} \right.$
\vspace{0.3 cm}

\noindent The function $Z$ is maximized if
$x_{2}^{+}=U, x_{2}^{-}=U$, and
$ x_{1}^{+} - x_{1}^{-} =U$. Thus
\begin{eqnarray} {\nonumber}
Z &\le&
U
\left(2y_{2}^{+} - A -V \right )
+U
\left(2y_{2}^{-} + A - V \right )
+U
\left(A + y_{2}^{+}-y_{2}^{-} \right ) \\ \nonumber
&-& U y_{2}^{+} -  U y_{2}^{-}
-UV \\
&=& -U\left(-2y_{2}^{+} - A + 3V \right ).
\end{eqnarray}
Since $V \ge -A$ and $V \ge y_{2}^{+}$, $Z \le 0$,
and the theorem is proved.

Now let $\mbox {\boldmath $a$ ($b$)}$
and $\mbox {\boldmath $a'$ ($b'$)}$
be two arbitrary orientation of the first
(second) polarizer, and let
\begin{eqnarray}{\nonumber}
x_{1}^{\pm}&=&p^{\;\pm}(\mbox{\boldmath $a$} \mid \lambda), \qquad
x_{2}^{\pm}=p^{\;\pm}(\mbox{\boldmath $a'$}|\lambda), \\ 
y_{1}^{\pm}&=&p^{\;\pm}(\mbox{\boldmath $b$}|\lambda), \qquad
y_{2}^{\pm}=p^{\;\pm}(\mbox{\boldmath $b'$}|\lambda).
\end{eqnarray}

\noindent Obviously for each value of $\lambda$, we have
\begin{eqnarray}{\nonumber}
p^{\;\pm}(\mbox{\boldmath $a$} \mid \lambda) \leq 1, \qquad
p^{\;\pm}(\mbox{\boldmath $a'$} \mid \lambda) \leq 1,\\ 
p^{\;\pm}(\mbox{\boldmath $b$} \mid \lambda) \leq 1, \qquad
p^{\;\pm}(\mbox{\boldmath $b'$} \mid \lambda) \leq 1.
\end{eqnarray}

\noindent Inequalities ($5$) and ($16$) yield
\begin{eqnarray} {\nonumber}
&&p^{+}(\mbox{\boldmath $a$} \mid \lambda) \,
p^{+}(\mbox{\boldmath $b$} \mid \lambda)
+p^{-}(\mbox{\boldmath $a$} \mid \lambda) \,
p^{-}(\mbox{\boldmath $b$} \mid \lambda)
-p^{+}(\mbox{\boldmath $a$} \mid \lambda) \,
p^{-}(\mbox{\boldmath $b$} \mid \lambda)  \\ \nonumber
&&- \,p^{-}(\mbox{\boldmath $a$} \mid \lambda) \, 
p^{+}(\mbox{\boldmath $b$} \mid \lambda)
+p^{+}(\mbox{\boldmath $b'$} \mid \lambda) \,
p^{+}(\mbox{\boldmath $a$} \mid \lambda) 
+p^{-}(\mbox{\boldmath $b'$} \mid \lambda) \,
p^{-}(\mbox{\boldmath $a$} \mid \lambda) \\ \nonumber
&&- \, p^{+}(\mbox{\boldmath $b'$} \mid \lambda) \,
p^{-}(\mbox{\boldmath $a$} \mid \lambda) -
p^{-}(\mbox{\boldmath $b'$} \mid \lambda) \,
p^{+}(\mbox{\boldmath $a$} \mid \lambda) -
p^{+}(\mbox{\boldmath $a'$} \mid \lambda) \,
p^{+}(\mbox{\boldmath $b$} \mid \lambda) \\ \nonumber
&&- \, p^{-}(\mbox{\boldmath $a'$} \mid \lambda) \,
p^{-}(\mbox{\boldmath $b$} \mid \lambda) +
p^{+}(\mbox{\boldmath $a'$} \mid \lambda) \,
p^{-}(\mbox{\boldmath $b$} \mid \lambda) 
+p^{-}(\mbox{\boldmath $a'$} \mid \lambda) \,
p^{+}(\mbox{\boldmath $b$} \mid \lambda) \\ \nonumber
&&+ \, 2p^{+}(\mbox{\boldmath $a'$} \mid \lambda) \, 
p^{+}(\mbox{\boldmath $b'$} \mid \lambda)+
2p^{-}(\mbox{\boldmath $a'$} \mid \lambda) \,
p^{-}(\mbox{\boldmath $b'$} \mid \lambda)- 
p^{+}(\mbox{\boldmath $a'$} \mid \lambda) \\
&&- \, p^{-}(\mbox{\boldmath $a'$} \mid \lambda)
-p^{+}(\mbox{\boldmath $b'$} \mid \lambda) \,
-p^{-}(\mbox{\boldmath $b'$} \mid \lambda) \le 1.
\end{eqnarray}

\noindent Multiplying both sides of $(17)$
by $p \, (\lambda)$, integrating over $\lambda$ and
using Eqs. $(3)$, we obtain
\begin{eqnarray} {\nonumber}
&&p^{+ +}(\mbox{\boldmath $a,\, b$}) 
+p^{- \,-}(\mbox{\boldmath $a,\, b$}) 
-p^{+ \,-}(\mbox{\boldmath $a,\, b$}) 
-p^{- \,+}(\mbox{\boldmath $a,\, b$})  
+p^{+ +}(\mbox{\boldmath $b',\, a$})+ \\ \nonumber
&&p^{- \,-}(\mbox{\boldmath $b',\, a$})
-p^{+ \,-}(\mbox{\boldmath $b',\, a$})
-p^{- \,+}(\mbox{\boldmath $b',\, a$}) 
-p^{+ +}(\mbox{\boldmath $a',\, b$})- \\ \nonumber
&&p^{- \,-}(\mbox{\boldmath $a',\, b$}) 
+p^{+ \,-}(\mbox{\boldmath $a',\, b$}) 
+p^{- \,+}(\mbox{\boldmath $a',\, b$}) 
+2p^{+ +}(\mbox{\boldmath $a',\, b'$}) + \\
&&2p^{- \,-}(\mbox{\boldmath $a',\, b'$}) 
-p^{+}(\mbox{\boldmath $a'$}) 
-p^{-}(\mbox{\boldmath $a'$}) 
-p^{+}(\mbox{\boldmath $b'$}) 
-p^{-}(\mbox{\boldmath $b'$})
\leq 1.
\end{eqnarray}
We now note that
the expected value of detection probabilities while polarizers
are set along orientations $\mbox{\boldmath $m$}$ and
$\mbox{\boldmath $n$}$, i.e.,
$E \,\left (\mbox{\boldmath $m,n$} \right)$
is defined as
\begin{eqnarray} \nonumber
E \,\left (\mbox{\boldmath $m,n$} \right) &= &
p^{+\, +} \,\left (\mbox{\boldmath $m,n$} \right)-
p^{+\, -} \,\left (\mbox{\boldmath $m,n$} \right) \\
&-&p^{-\, +} \,\left (\mbox{\boldmath $m,n$} \right)+
p^{-\, -} \,\left (\mbox{\boldmath $m,n$} \right).
\end{eqnarray}
Using $(19)$,
inequality $(18)$ may be written as
\begin{eqnarray} {\nonumber}
&&E(\mbox{\boldmath $a,\, b$}) +
E(\mbox{\boldmath $b',\, a$})
-E(\mbox{\boldmath $a',\, b$})
+2p^{+ +}(\mbox{\boldmath $a',\, b'$}) +
2p^{- \,-}(\mbox{\boldmath $a',\, b'$}) \\
&&-p^{+}(\mbox{\boldmath $a'$})
-p^{-}(\mbox{\boldmath $a'$})
-p^{+}(\mbox{\boldmath $b'$})
-p^{-}(\mbox{\boldmath $b'$})
\leq 1.
\end{eqnarray}
All local realistic theories must satisfy inequality $(18)$ or
$(20)$.

\begin{center}
\Large  {\bf IV. 
Violation of Bell's inequality in the  case of ideal experiments}
\end{center}

First we consider an
atomic cascade experiment in which
polarizers and detectors are ideal.
Assuming
polarizers are set along axes $\mbox{\boldmath $m$}$
and $\mbox{\boldmath $n$}$ where $\theta=
\mid \mbox{\boldmath $m$}-\mbox{\boldmath $n$} \mid$,
the expected values,
the single and joint detection
probabilities for a pair of photons in
a cascade from state $J=1$ to $J=0$
are given by
\begin{eqnarray} {\nonumber}
&&E \left ( \mbox{\boldmath $m,\, n$} \right)=
E \left(\theta \right) = \cos 2 \theta, \qquad
p^{+} (\mbox{\boldmath $a'$})=
p^{-} (\mbox{\boldmath $a'$})=
p^{+} (\mbox{\boldmath $b'$})=
p^{-} (\mbox{\boldmath $b'$})= \frac{1}{2}, \\ \nonumber
&&p^{+\, +} \left ( \mbox{\boldmath $m,\, n$} \right)=
p^{+\, +}\left(\theta \right) =
\frac{\cos^2 \theta}{2},
\qquad
p^{-\, -} \left ( \mbox{\boldmath $m,\, n$} \right)=
p^{-\, -} \left(\theta \right)
=\frac{\cos^2 \theta}{2}.
\\
\end{eqnarray}
Now if we choose the following orientation
$\left (\mbox{\boldmath $a , \, b$} \right)=
\left( \mbox{\boldmath $b' , \, a$} \right) =30^\circ$,
$\left (\mbox{\boldmath $a', \, b $} \right) = 60^\circ$
and
$ \left ( \mbox {\boldmath $a', \, b'$} \right) =0^\circ$
inequality $(20)$
becomes
\begin{eqnarray} \nonumber
&&2 E \left (30^\circ \right)
- E \left (60^\circ \right)
+2 p^{+\, +} \left (0^\circ \right)
+2 p^{-\, -} \left (0^\circ \right)-\\
&&p^{+} (\mbox{\boldmath $a'$})-
p^{-} (\mbox{\boldmath $a'$})-
p^{+} (\mbox{\boldmath $b'$})-
p^{-} (\mbox{\boldmath $b'$}) \le 1.
\end{eqnarray}
Using $(21)$, we obtain
\begin{eqnarray} {\nonumber}
&&2 \cos \left (60^\circ \right)
- \cos \left (120^\circ \right)
 +2 \frac{\cos^2 \left (0^\circ \right)}
{2}
+2 \frac{\cos^2 \left (0^\circ \right)}{2} - \frac{1}{2}
 - \frac{1}{2}
 - \frac{1}{2}
 - \frac{1}{2}\\
&&=2*(0.5)- (-0.5)+2*\frac{1}{2}+2*\frac{1}{2}-2 \le 1,
\end{eqnarray}
or
\begin{eqnarray}
1.5 \le 1,
\end{eqnarray}
which violates inequality $(20)$
by a factor of $1.5$
in the case of ideal experiments.

We now show that for ideal polarizers and detectors,
inequality $(20)$ is
equivalent to CHSH inequality.
In an ideal experiment, all emitted photons are analyzed
by the detectors
and the probability that a
photon is not collected is zero, i.e.,
\begin{eqnarray}
p^{\pm\, 0} \,\left ( \mbox{\boldmath $m, n$} \right)=
p^{0\, \pm} \,\left ( \mbox{\boldmath $m, n$} \right)=
p^{0\, 0} \,\left ( \mbox{\boldmath $m, n$} \right)=
p^{\, 0} \,\left ( \mbox{\boldmath $m$} \right)=
p^{\, 0} \,\left ( \mbox{\boldmath $n$} \right)=0
\end{eqnarray}
Thus in an ideal experiment,
\begin{eqnarray} \nonumber
&&p^{+} \,\left (\mbox{\boldmath $a'$} \right)=
p^{+\, +} \,\left ( \mbox{\boldmath $a', \, b'$} \right)+
p^{+\, -} \,\left ( \mbox{\boldmath $a', \, b'$} \right),
\\
&&p^{-} \,\left (\mbox{\boldmath $a'$} \right)=
p^{-\, +} \,\left ( \mbox{\boldmath $a', \, b'$} \right)+
p^{-\, -} \,\left ( \mbox{\boldmath $a', \, b'$} \right).
\end{eqnarray}
Substituting $(26)$ in $(20)$, we obtain
\begin{eqnarray} {\nonumber}
&&E(\mbox{\boldmath $a,\, b$}) +
E(\mbox{\boldmath $b',\, a$})
-E(\mbox{\boldmath $a',\, b$})+
p^{+\, +}(\mbox{\boldmath $a',\, b'$}) -
p^{+\, -}(\mbox{\boldmath $a',\, b'$}) - \\
&&p^{- \,+}(\mbox{\boldmath $a',\, b'$})+
p^{- \,-}(\mbox{\boldmath $a',\, b'$})
\leq 1 +p^{+} \,\left (\mbox{\boldmath $b'$} \right)
+p^{-} \,\left (\mbox{\boldmath $b'$} \right).
\end{eqnarray}
Since we have assumed detectors and polarizers are ideal, we have
$p^{+} \,\left (\mbox{\boldmath $b'$} \right)
+p^{-} \,\left (\mbox{\boldmath $b'$} \right)=1$.
Thus
\begin{eqnarray}
E(\mbox{\boldmath $a,\, b$}) +
E(\mbox{\boldmath $b',\, a$})
-E(\mbox{\boldmath $a',\, b$})+
E(\mbox{\boldmath $a',\, b'$})
\leq 2,
\end{eqnarray}
which is the same as CHSH inequality.
Thus for
ideal polarizers and detectors, the inequality derived in
this paper is equivalent to
CHSH inequality.

It is important to emphasize that in the case of ideal experiments,
neither the present
inequality nor CHSH are necessary: they both immediately
reduce to
Bell's original inequality of 1965 \cite{1}.
First we show that in ideal experiments, CHSH reduces to 
Bell's original inequality.
If we assume
$\mbox{\boldmath $a'$}$ and $\mbox{\boldmath $b'$}$
are along the same direction, using 
Eq. $(21)$,
we have
$E(\mbox{\boldmath $a',\, b'$})=1$.
CHSH inequality $(28)$ therefore
becomes
\begin{eqnarray}
E(\mbox{\boldmath $a,\, b$})+
E(\mbox{\boldmath $b',\, a$})-
E(\mbox{\boldmath $a',\, b$})
\le 1.
\end{eqnarray}
which is the same as Bell's original inequality of $1965$ \cite{1}.
If we choose the following orientations:
$\left (\mbox{\boldmath $a , \, b$} \right)=
\left( \mbox{\boldmath $b' , \, a$} \right) =30^\circ$,
$\left (\mbox{\boldmath $a', \, b $} \right) = 60^\circ$,
Bell's inequality $(29)$ is violated by a 
maximum factor of $1.5$.

We now show that
inequality $(20)$ also reduces to Bell's original inequality \cite{1}
in an ideal experiment.
Again if we assume
$\mbox{\boldmath $a'$}$ and $\mbox{\boldmath $b'$}$
are along the same direction, using $(21)$, we have
$p^{+\, +}(\mbox{\boldmath $a',\, b'$})=
p^{-\, -}(\mbox{\boldmath $a',\, b'$})=
\frac{\displaystyle 1}{\displaystyle 2}$,
$p^{+} \,\left (\mbox{\boldmath $a'$} \right)=
p^{-} \,\left (\mbox{\boldmath $a'$} \right)=
p^{+} \,\left (\mbox{\boldmath $b'$} \right)=
p^{-} \,\left (\mbox{\boldmath $b'$} \right)=
\frac{\displaystyle 1}{\displaystyle 2}$.
Inequality $(20)$ therefore
becomes
\begin{eqnarray}
E(\mbox{\boldmath $a,\, b$})+
E(\mbox{\boldmath $b',\, a$})-
E(\mbox{\boldmath $a',\, b$})
\le 1.
\end{eqnarray}
which is the same as Bell's original inequality of $1965$ \cite{1}.

We have thus shown that for
ideal polarizers and
detectors
Bell's original inequality $(29)$ is sufficient and
there is no need for inequality $(20)$ or CHSH inequality $(28)$.
Moreover, we have
shown that for the case ideal experiments (see $25$), i. e.,
for the case where
$p^{+\, +}(\mbox{\boldmath $m,\, n$})+
p^{+\, -}(\mbox{\boldmath $m,\, n$})+
p^{-\, +}(\mbox{\boldmath $m,\, n$})+
p^{+\, -}(\mbox{\boldmath $m,\, n$}) =1$,
inequality $(20)$ is equivalent to CHSH inequality and to Bell's 
original inequality.
However, for the case of
real experiments where
$p^{\pm\, 0}(\mbox{\boldmath $m,\, n$})$,
$p^{0\, \pm}(\mbox{\boldmath $m,\, n$})$, and
$p^{0\, 0}(\mbox{\boldmath $m,\, n$})$ are non-zero,
i. e., for the case where
$p^{+\, +}(\mbox{\boldmath $m,\, n$})+
p^{+\, -}(\mbox{\boldmath $m,\, n$})+
p^{-\, +}(\mbox{\boldmath $m,\, n$})+
p^{+\, -}(\mbox{\boldmath $m,\, n$}) <1$,
inequality $(20)$  is a distinct and new inequality
and is not equivalent to any of the previous inequalities.

\begin{center}
\Large  {\bf V. 
Violation of Bell's inequality in the case of real experiments}
\end{center}

We now consider a real
experiment in which
polarizers and detectors are non-ideal.
In the atomic cascade experiment, an atom emits two photons in 
a cascade from state $J=1$ to $J=0$. Since the pair of photons
have zero angular momentum, they propagate in the form of spherical
wave. Thus the probability $p \left(\mbox{\boldmath $d_1$},
\mbox{\boldmath $d_2$} \right)$ 
of both photons being simultaneously detected
by two detectors in the directions $\mbox{\boldmath $d_1$}$ and
$\mbox{\boldmath $d_2$}$ is  \cite{3},\cite{4}
\begin{eqnarray}
p \left(\mbox{\boldmath $d_1,\,d_2$} \right)=
\eta^2 \left ({\frac{\displaystyle \Omega}
{\displaystyle 4\pi}}\right) ^2
g \left (\theta,\phi \right ),
\end{eqnarray}
where $\eta$ is the quantum efficiency of the detectors, 
$\Omega$ is the solid angle of the detector, 
$\cos \theta=\mbox{\boldmath $d_1. d_1$}$,
and angle $\phi$ is related to $\Omega$ by
\begin{eqnarray}
\Omega=2 \pi \left (1-\cos \phi \right).
\end{eqnarray}
Finally the function 
$g \left (\theta,\phi \right )$ is the angular correlation function
and in the special case is given by \cite{4}
\begin{eqnarray} 
g \left (\pi, \phi \right ) = 1+
\frac{1}{8} \cos^2 \phi \left (1 + \cos \phi \right)^2.
\end{eqnarray}
If we insert polarizers in front of the detectors, then the
quantum mechanical predictions for
joint detection probabilities are \cite{3},
\cite{4}
\begin{eqnarray}  \nonumber
p^{+} \left ( \mbox{\boldmath $a$} \right )=
p^{-} \left ( \mbox{\boldmath $a$} \right )=
\eta \left ({\frac{\displaystyle \Omega}{\displaystyle 8 \pi}}
\right), \qquad
p^{+} \left ( \mbox{\boldmath $b$} \right )=
p^{-} \left ( \mbox{\boldmath $b$} \right )=
\eta \left ({\frac{\displaystyle \Omega}{\displaystyle 8 \pi}}
\right), \\ \nonumber
p^{+ \, +} \left ( \mbox{\boldmath $a,\, b$} \right )=
\eta^2 \left ({\frac{\displaystyle \Omega}{\displaystyle 8 \pi}}
\right)^2
g \left (\theta,\phi \right )
\left[T_+^1T_+^2+ T_-^1T_-^2  
F (\theta)
\cos 2 \left ( \mbox{\boldmath $a- b$} \right ) \right ], \\  \nonumber
p^{- \, -} \left ( \mbox{\boldmath $a,\, b$} \right )=
\eta^2 \left ({\frac{\displaystyle \Omega}{\displaystyle 8 \pi}}
\right)^2
g \left (\theta,\phi \right )
\left[ R_+^1R_+^2+ R_-^1R_-^2  
F (\theta) 
\cos 2 \left ( \mbox{\boldmath $a- b$} \right ) \right ], \\  \nonumber
p^{+ \, -} \left ( \mbox{\boldmath $a,\, b$} \right )=
\eta^2 \left ({\frac{\displaystyle \Omega}
{\displaystyle 8 \pi}}\right)^2
g \left (\theta,\phi \right )
\left[T_+^1R_+^2- T_-^1R_-^2 
F (\theta) \cos 2 \left ( \mbox{\boldmath $a- b$} \right ) \right],
\\ 
p^{- \, +} \left ( \mbox{\boldmath $a,\, b$} \right )=
\eta^2 \left ({\frac{\displaystyle \Omega}
{\displaystyle 8 \pi}}\right)^2
g \left (\theta,\phi \right )
\left[R_+^1T_+^2- R_-^1T_-^2 
F (\theta) \cos 2\left ( \mbox{\boldmath $a- b$} \right ) \right),
\end {eqnarray}
where
\begin {eqnarray}
&&T_+^i=T_{\|}^i+T_{\perp}^i, \qquad
T_-^i=T_{\|}^i-T_{\perp}^i \\ \nonumber
&&R_+^i=R_{\|}^i+R_{\perp}^i, \qquad
R_-^i=R_{\|}^i-R_{\perp}^i 
\end {eqnarray}
for $i=1,2$, where $T_{\|}(T_{\perp})$ represents the 
prism transmittance along the transmitted path
for incoming light polarized parallel (perpendicular) to the 
transmitted-channel, and
$R_{\|}(R_{\perp})$ represents the 
prism transmittance along the reflected path
for incoming light polarized parallel (perpendicular) to the 
reflected-channel.
The function $F \left (\theta,\phi \right )$
 is the so-called depolarization
factor and for the special case $\theta=\pi$ and small $\phi$
is given by
\begin{eqnarray}
F (\pi, \phi) \approx 1- \frac{2}{3} \left (1-\cos \phi
\right)^{2}.
\end{eqnarray}
The function $F \left (\theta,\phi \right )$,
in general, is very close to $1$ (the detailed expression for
$F \left (\theta,\phi \right )$ is given in \cite {4}).

In the atomic cascade
experiments which are feasible with present technology [5,12],
because
$\frac{\displaystyle \Omega}{\displaystyle  4 \pi} \ll 1$,
only a very small fraction of photons are detected.
Thus inequality
$(18)$ can not be used to test the violation of Bell's 
inequality. 
It is important to emphasize that
a supplementary assumption is required primarily
because the solid angle covered by the
aperture of the apparatus,
$\Omega$, is  much less than $4 \pi$ and not because the
efficiency of the detectors, $\eta$, is much smaller than $1$. In
fact in the previous experiments (Ref. $12$),
the efficiency of detectors were larger than
$90 \%$.
However, because
$\frac{\displaystyle \Omega}{\displaystyle  4 \pi} \ll 1$,
all previous experiments needed
supplementary assumptions to test locality.

It is worth nothing that
CHSH \cite{4} and CH \cite{6} combined the solid angle covered
by the aperture of the apparatus 
$\frac{\displaystyle \Omega}{\displaystyle  4 \pi}$ and the
efficiency of the detectors $\eta$ into one term and wrote
$\eta_{CHSH}=\eta \frac 
{\displaystyle \Omega}{\displaystyle 4 \pi}$. They then referred to
$\eta_{CHSH}$ as efficiency of the detectors (this terminology 
however, is not usually used in optics. In optics, the efficiency of 
detector refers to the probability of
detection of a photon; it does not refer to the product of 
the solid angle covered by the detector and the
probability of detection of a photon).
CHSH and CH then pointed that
since $\eta_{CHSH} \ll 1$,
a supplementary assumption is required. To clarify
CHSH and CH argument, it should be emphasized that a supplementary
assumption is needed mainly because 
$\frac{\displaystyle \Omega}{\displaystyle  4 \pi} \ll 1$, not
because $\eta \ll 1$.

We now
state a supplementary assumption and we
show that this assumption is sufficient to make
experiments where $\frac{\displaystyle \Omega}
{\displaystyle  4 \pi} \ll 1$
applicable as a test of local theories.
The supplementary assumption is:
For every emission $\lambda$, the detection probability 
by detector $D^{+}$ (or $D^-$) 
is {\it less than or equal} to the sum  of detection probabilities
by detectors $D^{+}$ and $D^-$ when the polarizer
is set along any {\it arbitrary} axis.
If we let $\mbox{\boldmath $r$}$ be an {\it arbitrary} direction
of the 
first or second polarizer,
then the above supplementary assumption
may be translated into the 
following inequalities
\begin{eqnarray} {\nonumber}
p^{\;+}(\mbox{\boldmath $a$} \mid \lambda) \leq 
p^{\;+}(\mbox{\boldmath $r$} \mid \lambda)+
p^{\;-}(\mbox{\boldmath $r$} \mid \lambda), \qquad
p^{\;-}(\mbox{\boldmath $a$} \mid \lambda) \leq 
p^{\;+}(\mbox{\boldmath $r$} \mid \lambda)+
p^{\;-}(\mbox{\boldmath $r$} \mid \lambda), \\ \nonumber
p^{\;+}(\mbox{\boldmath $a'$} \mid \lambda) \leq 
p^{\;+}(\mbox{\boldmath $r$} \mid \lambda)+
p^{\;-}(\mbox{\boldmath $r$} \mid \lambda), \qquad
p^{\;-}(\mbox{\boldmath $a'$} \mid \lambda) \leq 
p^{\;+}(\mbox{\boldmath $r$} \mid \lambda)+
p^{\;-}(\mbox{\boldmath $r$} \mid \lambda), \\ \nonumber
p^{\;+}(\mbox{\boldmath $b$} \mid \lambda) \leq 
p^{\;+}(\mbox{\boldmath $r$} \mid \lambda)+
p^{\;-}(\mbox{\boldmath $r$} \mid \lambda), \qquad
p^{\;-}(\mbox{\boldmath $b$} \mid \lambda) \leq 
p^{\;+}(\mbox{\boldmath $r$} \mid \lambda)+
p^{\;-}(\mbox{\boldmath $r$} \mid \lambda), \\  \nonumber
p^{\;+}(\mbox{\boldmath $b'$} \mid \lambda) \leq 
p^{\;+}(\mbox{\boldmath $r$} \mid \lambda)+
p^{\;-}(\mbox{\boldmath $r$} \mid \lambda), \qquad
p^{\;-}(\mbox{\boldmath $b'$} \mid \lambda) \leq 
p^{\;+}(\mbox{\boldmath $r$} \mid \lambda)+
p^{\;-}(\mbox{\boldmath $r$} \mid \lambda).
\\
\end{eqnarray}
This supplementary assumption is obviously valid for an ensemble
of photons. 
The sum of detection probability
by detector $D^{+}$ and $D^-$
for an ensemble of photons
when the polarizer
is set along any {\it arbitrary} axis $\mbox{\boldmath $v$}$
\begin{eqnarray}
p^{+}(\mbox{\boldmath $v$})+p^{-}(\mbox{\boldmath $v$})=
\eta \left ({\frac{\displaystyle \Omega}{\displaystyle 4 \pi}} \right),
\end{eqnarray} 
whereas 
\begin{eqnarray} 
p^{+}(\mbox{\boldmath $v$})=p^{-}(\mbox{\boldmath $v$})=
\eta \left ({\frac{\displaystyle \Omega}{\displaystyle 8 \pi}} \right).
\end{eqnarray} 
The supplementary assumption requires that the corresponding
probabilities
be valid for each $\lambda$.

It is worth noting that the present supplementary assumption is
weaker than Garuccio-Rapisarda (GR) assumption \cite{8},
that is, an experiment based on the present supplementary assumption
refutes a larger family of hidden variable theories than an 
experiment based on GR assumption.
The GR assumption is
\begin{eqnarray} 
p^{+}(\mbox{\boldmath $a$} \mid \lambda)+
p^{-}(\mbox{\boldmath $a$} \mid \lambda)=
p^{+}(\mbox{\boldmath $r$} \mid \lambda)+
p^{-}(\mbox{\boldmath $r$} \mid \lambda)
\end{eqnarray} 
We now show that 
GR assumption implies the 
assumption of this paper.
We first note that the following inequalities always
hold
\begin{eqnarray} \nonumber
p^{+}(\mbox{\boldmath $a$} \mid \lambda) \le
p^{+}(\mbox{\boldmath $a$} \mid \lambda) +
p^{-}(\mbox{\boldmath $a$} \mid \lambda), \qquad
p^{-}(\mbox{\boldmath $a$} \mid \lambda) \le
p^{+}(\mbox{\boldmath $a$} \mid \lambda) +
p^{-}(\mbox{\boldmath $a$} \mid \lambda). \\
\end{eqnarray} 
Now using GR assumption $(40)$, 
we can immediately conclude that
\begin{eqnarray} \nonumber
p^{+}(\mbox{\boldmath $a$} \mid \lambda) \le
p^{+}(\mbox{\boldmath $r$} \mid \lambda)+
p^{-}(\mbox{\boldmath $r$} \mid \lambda), \qquad
p^{-}(\mbox{\boldmath $a$} \mid \lambda) \le
p^{+}(\mbox{\boldmath $r$} \mid \lambda)+
p^{-}(\mbox{\boldmath $r$} \mid \lambda), \\
\end{eqnarray} 
which are the same as $(37)$.
Thus an
experiment which refutes the
hidden variable theories which are consistent with GR assumption also
refutes the hidden variable theories which are consistent with the
present assumption. The reverse however is not true. An experiment
based on the present supplementary assumption refutes a larger
family of hidden variable theories than an experiment based on GR
assumption. 

Now using relations $(5)$, $(15)$ and $(37)$, and applying
the same argument that
led to inequality $(18)$, we obtain the following inequality
\begin{eqnarray} {\nonumber}
&&\bigg[p^{+ \,+}(\mbox{\boldmath $a,\, b$})
+p^{- \,-}(\mbox{\boldmath $a,\, b$})
-p^{+ \,-}(\mbox{\boldmath $a,\, b$})
-p^{- \,+}(\mbox{\boldmath $a,\, b$}) 
+p^{+ +}(\mbox{\boldmath $b',\, a$})   
+p^{- \,-}(\mbox{\boldmath $b',\, a$})   \\  \nonumber
&&-p^{+ \,-}(\mbox{\boldmath $b',\, a$}) 
-p^{- \,+}(\mbox{\boldmath $b',\, a$}) 
-p^{+ \,+}(\mbox{\boldmath $a',\, b$}) 
-p^{- \,-}(\mbox{\boldmath $a',\, b$})
+p^{+ \,-}(\mbox{\boldmath $a',\, b$}) \\  \nonumber
&&+p^{- \,+}(\mbox{\boldmath $a',\, b$}) 
+2p^{+ +}(\mbox{\boldmath $a',\, b'$})
+2p^{- \,-}(\mbox{\boldmath $a',\, b'$})
-p^{+\,+}(\mbox{\boldmath $a',r$})
-p^{+\,-}(\mbox{\boldmath $a',r$}) \\  \nonumber
&&-p^{-\,+}(\mbox{\boldmath $a',r$}) 
-p^{-\,-}(\mbox{\boldmath $a',r$}) 
-p^{+\,+}(\mbox{\boldmath $r,b'$})
-p^{+\,-}(\mbox{\boldmath $r,b'$})
-p^{-\, +}(\mbox{\boldmath $r,b'$})\\  
&&-p^{-\, -}(\mbox{\boldmath $r,b'$}) \bigg ] \, \bigg / \,
\left[p^{+\,+}(\mbox{\boldmath $r,r$}) 
+p^{+\,-}(\mbox{\boldmath $r,r$})
+p^{-\,+}(\mbox{\boldmath $r,r$})
+p^{-\,-}(\mbox{\boldmath $r,r$}) \right ] \leq 1.
\end{eqnarray}
\noindent Note that
inequality $(43)$ contains only double-detection
probabilities and
the number of emissions $N$ from the source
is eliminated from the ratio.
Quantum mechanics violates
this inequality
in the case of real experiments where the solid angle covered 
by the aperture of the apparatus, $\Omega$, is  much less than
$4 \pi$. In particular, the magnitude of violation is maximized if the
following orientations are chosen
$\left (\mbox{\boldmath $a , \, b$} \right)=
\left( \mbox{\boldmath $b' , \, a$} \right)=30^\circ$,
$\left( \mbox{\boldmath $a' , \, b$} \right)= 60^\circ$
and
$ \left( \mbox {\boldmath $a', \, b'$} \right)=
\left ( \mbox {\boldmath $a', \, r$} \right) =
\left ( \mbox {\boldmath $r , \, b'$} \right) = 0^\circ$.

Inequality $(43)$ may be considerably simplified if we invoke some
of the symmetries that are exhibited in atomic-cascade photon
experiments. For a pair of photons in
cascade from state $J=1$ to $J=0$, the
quantum mechanical detection probabilities $p^{\pm\, \pm}_{QM}$ and
expected value $E_{QM}$ exhibit the following
symmetry
\begin{eqnarray} 
p^{\pm\, \pm}_{QM} \,\left (\mbox{\boldmath $m,n$} \right)=
p^{\pm\, \pm}_{QM}
\,\left ( \mid\mbox{\boldmath $m-n$} \mid\right), \qquad
E_{QM} \,\left (\mbox{\boldmath $m,n$} \right)=
E_{QM} \,\left ( \mid\mbox{\boldmath $m-n$} \mid\right).
\end{eqnarray}
We assume that the 
local theories also exhibit the same symmetry
\begin{eqnarray}
p^{\pm\, \pm} \,\left (\mbox{\boldmath $m,n$} \right)=
p^{\pm\, \pm} \,\left ( \mid\mbox{\boldmath $m-n$} \mid\right), \qquad
E \,\left (\mbox{\boldmath $m,n$} \right)=
E \,\left ( \mid\mbox{\boldmath $m-n$} \mid\right).
\end{eqnarray}
Note that there is no harm in assuming Eqs.
$(45)$ since they are subject to experimental test (CHSH
\cite {4}, FC \cite{5}, and CH \cite {6}
made the same assumptions).
Using the above symmetry, inequality $(43)$ is simplified to
\begin{eqnarray} \nonumber
&&\bigg [E \,\left (\mid \mbox{\boldmath $a-b$} \mid \right)+
E \,\left (\mid \mbox{\boldmath $b'-a$} \mid \right)-
E \,\left (\mid \mbox{\boldmath $a'-b$} \mid \right)+
2p^{+\, +} \,\left (\mid\mbox{\boldmath $a'-b'$} \mid\right)
+2p^{-\, -} \,\left (\mid\mbox{\boldmath $a'-b'$} \mid\right) \\ 
\nonumber
&&-p^{+\, +} \,\left (\mid\mbox{\boldmath $a'-r$} \mid\right)-
p^{+\, -} \,\left (\mid\mbox{\boldmath $a'-r$} \mid\right)-
p^{-\, +} \,\left (\mid\mbox{\boldmath $a'-r$} \mid\right) 
-p^{-\, -} \,\left (\mid\mbox{\boldmath $a'-r$} \mid\right)\\
\nonumber
&&-p^{+\, +} \,\left (\mid\mbox{\boldmath $r-b'$} \mid\right)-
p^{+\, -} \,\left (\mid\mbox{\boldmath $r-b'$} \mid\right)-
p^{-\, +} \,\left (\mid\mbox{\boldmath $r-b'$} \mid\right)
-p^{-\, -} \,\left (\mid\mbox{\boldmath $r-b'$} \mid\right)
\bigg ] \, \bigg / \, \\ 
&&\bigg [p^{+\,+}\, (0^\circ) +
p^{+\,-}\, (0^\circ) +
p^{-\,+}\, (0^\circ) +
p^{-\,-}\, (0^\circ) 
\bigg ] \leq 1.
\end{eqnarray}
We now take $\mbox{\boldmath $a'$}$ and 
$\mbox{\boldmath $b'$}$ to be along
$\mbox{\boldmath $r$}$,
and we take
$\mbox{\boldmath $a$}$, 
$\mbox{\boldmath $b$}$, and $\mbox{\boldmath $a'$}$ to be
three coplanar axes
with the following orientations:
$\mid \mbox{\boldmath $a - b$} \mid=
 \mid \mbox{\boldmath $b' - a$} \mid= 30^\circ$,
 $\mid \mbox{\boldmath $a' - b$} \mid= 60^\circ $
and 
$ \mid \mbox {\boldmath $a'- b'$} \mid= 
\mid \mbox {\boldmath $a'- r$} \mid= 
\mid \mbox {\boldmath $r- b'$} \mid= 0^\circ $.
Furthermore if we define $K$ as
\begin{eqnarray}
K=p^{+\,+}(0^\circ)
+p^{+\,-}(0^\circ)
+p^{-\,+}(0^\circ)
+p^{-\,-}(0^\circ)
\end{eqnarray}
then inequality $(46)$ is simplified to
\begin{eqnarray}
\frac {2E \left( 30^\circ \right)-
E \left( 60^\circ \right)
-2p^{+\, -}\left( 0^\circ \right)
-2p^{- \,+}\left( 0^\circ \right)}{K} \leq 1.
\end{eqnarray}
Using the quantum mechanical probabilities
[i.e., Eqs. $(34)$],
inequality $(48)$ becomes
$1.5 \leq 1$ in the case of real experiments
(here for simplicity, we have assumed
$F \left (\theta,\phi \right ) =1$; this is a good approximation
even in the case of real experiments. In
actual experiments where the solid angle of detectors
$\phi$ is usually less than $\pi/6$, from $(36)$ it can be seen that
$F (\theta, \pi/6) \approx 0.99$. Moreover, we have assumed 
$T_{\|}^i=R_{\|}^i=1$ and 
$T_{\perp}^i=R_{\perp}^i=0$, 
where $i=1,2$; this is also a good approximation,
see for example \cite{3} or the experiments by Aspect {\em et. al.}
\cite{12}).

Inequality $(48)$ can be used to test locality 
more simply than CH or CHSH inequality.
CH inequality may 
be written as
\begin{eqnarray}
\frac{3 p \left( \phi \right)- 
p \left( 3 \phi \right)- 
p \left( \mbox{\boldmath $a'$}, \infty \right)- 
p \left( \infty , \mbox{\boldmath $b$} \right)}{
p \left( \infty , \infty \right)} \leq 0.
\end{eqnarray}
The above inequality requires the measurements of five
detection probabilities:
\\
(1) The measurement of detection probability
with both polarizers set along the $22.5^\circ$ axis
[that is $p\left(22.5^\circ \right)$].
\\
(2) The measurement of detection probability
with both polarizers set along the $67.5^\circ$ axis
[that is $p\left(67.5^\circ \right)$].
\\
(3) The measurement of detection probability
with the first polarizer set along
$\mbox{\boldmath $a'$}$ axis and the second polarizer being
removed
[that is 
$ p\left(\mbox{\boldmath $a'$}, \infty \right)$].
\\
(4) The measurement of detection probability
with the first polarizer removed and the second polarizer set along
$\mbox{\boldmath $b$}$ axis
[that is 
$ p\left(\infty , \mbox{\boldmath $b$} \right)$].
\\
(5) The measurement of detection probability
with both polarizers removed [that is 
$ p \left( \infty , \infty \right)$].
\\
In contrast, the inequality derived in this paper
[i.e., inequality $(48)$] requires the measurements of only
three detection probabilities:
\\
(1) The measurement of detection probability
with both polarizers set along the $0^\circ$ axis
[that is $p^{\pm \, \pm}\left(0^\circ \right)$].
\\
(2) The measurement of detection probability
with both polarizers set along the $30^\circ$ axis
[that is $p^{\pm \, \pm}\left(30^\circ \right)$].
\\
(3) The measurement of detection probability
with both polarizers set along the $60^\circ$ axis
[that is $p^{\pm \, \pm}\left(60^\circ \right)$].
\begin{center}
\Large  {\bf VI. Violation of Bell's inequality in 
phase-momentum and in high-energy experiments}
\end{center}

It should be noted that the analysis that led to inequality $(48)$
is not limited to atomic-cascade experiments and
can easily be extended to experiments which use
phase-momentum \cite {13} or use high energy
polarized protons or $\gamma$ photons [14-15] to test Bell's limit.
For example in the experiment by Rarity and Tapster \cite{13},
instead of inequality $(2)$ of their paper, the following
inequality (i.e., inequality $(48)$ 
using their notations) may be used to test locality:
\begin{eqnarray}
\frac {2E \left (30^\circ \right)
-E \left (60^\circ \right)
-2{\overline C}_{a_3 \,b_4}\left( 0^\circ \right)
-2{\overline C}_{a_4 \,b_3}\left( 0^\circ \right)}
{K} \leq 1,
\end{eqnarray}
where ${\overline C} _{a_i \,b_j} \left ( \phi_a \, , \phi_b \right)$
$(i=3,4;j=3,4)$ is the counting rate between detectors $D_{ai}$ and
$D_{bj}$ with phase angles being 
set to $\phi_a \, , \phi_b$ (See Fig. 1 of \cite {13}).
The following set of orientations
$\mbox{\boldmath $(\phi_a,\, \phi_b)$}=
\mbox{\boldmath $(\phi_{b'},\, \phi_a)$}= 30^\circ $, 
$\mbox{\boldmath $(\phi_{a'},\, \phi_{b})$}=60^\circ $, and
$\mbox{\boldmath $(\phi_{a'},\, \phi_{b'})$}=0^\circ $
leads to the largest violation.
Similarly, in high-energy experiments and
spin correlation proton-proton scattering experiments \cite {15},
inequality $(48)$ can be used to test locality.

\begin{center}
\Large  {\bf VII. Summary}
\end{center}

We have invoked Einstein's locality (Eq. $4$) to derive
a correlation inequality [inequality $(20)$] that can be used to test
locality.
In the case of ideal
experiments, this inequality is equivalent to Bell's
original inequality of $1965$ \cite{1}
or CHSH inequality \cite{4}. However, in the 
case of real experiments where
polarizers and detectors
are non-ideal, inequality $(20)$ is a new and distinct inequality and
is not equivalent to any of
previous inequalities.

We have also demonstrated that
the conjunction of Einstein's locality
[Eq. $(4)$] with a supplementary assumption [inequality $(37)$]
leads to validity of inequality $(48)$
that is sometimes grossly violated
by quantum mechanics in the case of experiments where
$\frac{\displaystyle \Omega}{\displaystyle  4 \pi} \ll 1$.
Inequality $(48)$, which may be called 
{\em strong} inequality \cite{16},
defines an experiment which can actually
be performed with present technology.
Quantum mechanics violates 
inequality
$(48)$ by a factor $1.5$, whereas it violates CHSH or CH inequality by 
a factor of $\sqrt 2$.
Thus the magnitude of violation of the inequality derived in this
paper is
approximately $20.7\%$ larger than the magnitude of violation of
previous inequalities [4-10].
The larger magnitude of violation 
can be useful for experimental test of locality.

\pagebreak

\end{document}